\documentstyle[11pt,psfig,twoside]{article}
\begin{document}\hbadness=10000
%
%
\pagestyle{myheadings}
\thispagestyle{empty}
\markboth{J. Letessier and  J. Rafelski}
{Chemical non-equilibrium and deconfinement 
in 200 A GeV Sulphur induced reactions}
\title{Chemical non-equilibrium and deconfinement\\ 
in 200 A GeV Sulphur induced reactions}
\author{Jean Letessier and Johann Rafelski\\ $ $\\
%
Laboratoire de Physique Th\'eorique et Hautes Energies\\
Universit\'e Paris 7, 2 place Jussieu, F--75251 Cedex 05.\\ 
and\\
Department of Physics, University of Arizona, Tucson, AZ 85721
}
\date{June 16, 1998; October 5, 1998}
\maketitle
%
\hspace*{0.4cm}\begin{minipage}[t]{15.5cm}\small\vskip -0.5cm
We interpret  hadronic particle abundances produced in 
S--Au/W/Pb 200 A GeV reactions in terms of the final state 
hadronic phase space model
and determine by a data fit of the chemical hadron
freeze-out parameters.  Allowing for the flavor abundance
non-equilibrium a 
highly significant fit to experimental particle abundance
data emerges, which supports possibility of 
strangeness distillation. We find
under different strategies stable values for freeze-out
temperature $T_{\rm f}=143\pm3$ MeV, baryochemical potential
$\mu_{\rm B}= 173\pm6$ MeV, ratio of strangeness 
($\gamma_{\rm s}$) and light quark ($\gamma_{\rm q}$)
phase space occupancies 
$\gamma_{\rm s}/\gamma_{\rm q}=0.60\pm0.02$, and 
$\gamma_{\rm q}=1.22\pm0.05$ without accounting for collective
expansion (radial flow). When introducing flow effects which
allow a consistent description of the transverse mass particle spectra,
yielding $|\vec v_{\rm c}|=0.49\pm0.01\,c$,
we find  $\gamma_{\rm s}/\gamma_{\rm q}=0.69\pm0.03$\,,
$\gamma_{\rm q}=1.41\pm0.08$. 
The strange quark fugacity is fitted at
$\lambda_{\rm s}=1.00\pm0.02$ suggesting chemical
freeze-out directly from  the deconfined  phase.
\end{minipage}
%
\section{Introduction}
In relativistic nuclear collisions, for a short period of time,
physical conditions are recreated similar to those present in
the early Universe about 40$\,\mu$s after the Big-Bang. 
The  primary objective of the extensive 
experimental program  at BNL (Brookhaven National Laboratory)
and CERN (European Organization for Nuclear Research) 
is the understanding of  `vacuum melting', 
the  formation and properties of the deconfined 
quark-gluon plasma (QGP) phase of strongly interacting (hadronic)
matter \cite{HM96}.

Our present investigation addresses the completed experimental 
analysis of the CERN 200 A GeV
Sulphur beam reactions with laboratory stationary 
`heavy' targets, such as Gold, Tungsten or Lead nuclei
This highest presently available energy  content 
materializes in form of  hadronic particle multiplicity
and corresponds to nuclear 
collisions occurring in the center of momentum frame 
$E_{\rm CM}=\sqrt{s}/B=8.8\,{\rm GeV}=9.4m_{\rm N}c^2$
for each participating  nucleon. 
Several prior attempts have been made to interpret
the hadronic particle abundances and spectra within 
statistical particle production interpretation
\cite{Raf91,LTR92,SCRS93,Let95,BSWX96,Sol97,GS98,BGS98},
and the present report takes, as we believe it, to a  conclusion
this extensive body of research work. 
The result of our present analysis are the parameters 
and properties of the hadronic system at chemical freeze-out. The
new theoretical element which makes this attempt successful and which we 
introduce here is the chemical non-equilibrium abundance of light 
quark flavors.

This analysis of the experimental data  is based
on the assumption that a relatively small and dense 
volume of highly excited hadronic matter, the `fireball', is
formed in the reaction. When we speak of
particle abundance freeze-out, we refer to the stage in evolution of the
fireball at which density has dropped to the level that
in subsequent collisions particle abundances remain unchanged.
We side-step  the need to understand largely  unknown collective 
flows originating in both, the 
memory of the initial `longitudinal' collision momentum, and 
the explosive `transverse' disintegration driven by the 
internal pressure of the hadronic matter. Thus, all
particles originate from an expanding fireball surface. 
In order to minimize the dependence on the source shape,
geometry and dynamics, we either consider ratios of particles obtained
at central rapidity, or ratios of $4\pi$-abundances.  
Suggestions were made that detailed production dynamics needs to be
 considered \cite{GS98},
allowing that  some particles are emitted well ahead of others, and 
of the final state hadronization. In our approach we see evidence for 
at best, two stage process, with some distillation of flavor
content, and specifically strangeness leading to $s$-quark enriched 
final stage. However, there is no compelling evidence for a real 
multi-stage, continuous 
particle production process in our evaluation of the data. 

\section{Chemical non-equilibrium statistical model}\label{secnoneq}
Along with the other statistical analysis of the 
experimental data mentioned above,
we employ the local thermal 
equilibrium method, and use a local freeze-out
temperature $T_{\rm f}$. One can argue that the accessibility
of many degrees of freedom, as expressed by high specific entropy
content, validates use of the thermal equilibrium. The usefulness 
of the local thermal equilibrium is, in part, 
based on the experimental fact that, at given transverse mass
$m_\bot=\sqrt{m^2+p_\bot^2}\simeq 1.5$--$2.5$ GeV, $m_\bot$-spectra of
very different hadronic particles have the 
same inverse slope $T_0$  \cite{acta96}.  To a very adequate
precision, for these high $m_\bot$, this inverse slope $T_0\simeq 235$ MeV
arises from $T_{\rm f}$  according to the  Doppler-blue-shift relation
using a local collective flow velocity. Impact  of
flow effects is discussed in detail
 in section \ref{secsum}.

Regarding chemical equilibration, we will  
recognize here two different types, relative and absolute
\cite{KMR86}, illustrated in the following example. Consider 
within a fixed volume a hot, 
thermally equilibrated gas of nucleons $N=p,n$, pions 
$\pi^\pm,\pi^0$, and $\Delta$-resonances: \\
$\bullet$ {\it Relative chemical equilibrium}\\
The relative abundance of the $u$ and $d$ carrying quarks is easily 
established through flavor exchange reactions, such as $p+\pi^-
\leftrightarrow n+\pi^0$; at the valence quark level there is no creation
or annihilation process that need to occur in each reaction to 
equilibrate, between the different hadronic particles, the quark 
flavors. We speak of relative chemical equilibration occurring 
through quark exchange reactions described in terms of 
fugacities $\lambda_i,\,i=u,\,d,\,s$. From quark fugacities, the 
particle fugacities are reconstituted, and in particular we recall
that baryochemical potential arises as $\mu_{\rm B}=3T\ln\lambda_{\rm q}$,
where we assumed that the difference between $u$ and $d$ quarks 
is not significantly affecting the particle ratios studied 
here \cite{Let95}. \\
$\bullet$ {\it Absolute chemical equilibrium}\\
Very much different is the
process involving the equilibration between the number densities of 
mesons and baryons.  Baryon-antibaryon formation 
processes are of the type
$N+\overline{N}\leftrightarrow\rho+\omega$ and require that aside of the 
reorganization of the quark content, also the number of valance quarks
changes, here between 6 and 4. The  valance quark numbers
 has to be controlled  by new parameters, the phase space occupancies
 $\gamma_i,\,i=u,\,d,\,s$. Only when $\gamma_i$ is present can
the freeze-out abundance of hadron families be
differentiated. 

The difference between $\lambda_i$ and $\gamma_i$ is that, {\it e.g.}, 
for strange and anti-strange quarks 
the same factor $\gamma_{\rm s}$ 
applies, while the antiparticle fugacity is inverse of the particle
fugacity. The proper statistical physics foundation of  $\gamma_i$ 
is obtained considering the maximum entropy principle 
for evolution of physical systems. In such a study it has
been determined that while the limit $\gamma_i\to1$ maximizes the specific
chemical entropy, this maximum is extremely shallow, indicating that 
a system with dynamically evolving volume will in general find more effective
 paths to increase entropy, than offered by the establishment of 
the absolute chemical equilibrium \cite{entro}. 

Generally, the microscopic processes which lead to absolute chemical
equilibrium  are  slower than those
leading to establishment of thermal or 
relative chemical equilibrium, considering that
the inelastic reactions that change the particle abundance have
usually much smaller cross section than particle number preserving 
exchange reaction, or elastic collisions. While thermal, 
and relative chemical equilibrium 
will in general occur within the life span of dense matter
formed in nuclear collisions, 
approach to absolute chemical equilibrium provides interesting 
chronometric and structure information about the dynamics of the 
collision  process \cite{KMR86,RM82}. In consequence, the number density 
of different hadrons produced in the collision 
has to be characterized by individual
fugacities. If each hadronic species were to require its
own fugacity, there would be little opportunity to relate reliably
the observed particle yields to specific physical processes. 

The interesting  insight
of this paper, concluded from the success of the description of the 
experimental data, is that chemical non-equilibrium is important
for the understanding of the experimental results, and that 
there is need for just  four chemical abundance parameters: 
two standard fugacities $\lambda_{\rm q}$ and $\lambda_{\rm s}$ 
(relative chemical equilibrium), and two chemical 
non-equilibrium occupancy parameters, measuring the
abundance of valance quarks, $\gamma_{\rm s}$, a parameter
describing the approach of strange quarks to phase space
abundance  equilibrium \cite{Raf91,KMR86,RM82}, and $\gamma_{\rm q}$,
introduced here for the first time into such data analysis,
characterizing the phase space abundance of light quarks. 
$\gamma_{\rm q}$ is  required as a parameter in the picture of
direct hadronization of a deconfined QGP 
region into final state hadrons, which are free 
streaming and do not form an intermediate hadronic phase. 
When developing  a dynamical description of QGP evaporation, one immediately 
establishes that the entropy excess we reported previously \cite{Let93},
requires a hadronic phase space parameter capable to control 
the excess of  hadronic  particle abundances consistently. 
Moreover, considering that in the fragmentation of gluons
additional light quark pairs arise, and can over populate the phase space abundance
of light quarks, we are naturally lead to implement the light quark occupancy 
parameter $\gamma_{\rm q}$. Note that such direct emission from QGP scenario, 
while convenient to understand the meaning of our data analysis in microscopic 
terms, is by far not a unique framework for the interpretation of the results 
we obtain. Which microscopic model is appropriate will be
understood when other systems can be subject to a similar analysis as 
the one presented here. Comparable particle abundance data is presently
assembled both for Au--Au collisions at 11 A GeV at AGS-BNL, and for
Pb--Pb collisions at 158~A GeV at CERN, with intermediate energies available
at CERN in near future. 

The evaluation of the final particle yields follows the pattern 
established in earlier work (see, {\it e.g.}, \cite{LTR92,Let95,acta96}).  
The relative number of primary particles 
freezing out from a source is obtained 
noting that the  fugacity and phase space occupancy 
 of a composite hadronic  particle is  expressed  
by its constituents and that the probability to find all 
$j$-components contained within  the $i$-th  emitted particle is:
\begin{equation}\label{abund}
N_i\propto \prod_{j\in i}\gamma_j\lambda_je^{-E_j/T}\,,
\qquad\lambda_i=\prod_{j\in i}\lambda_j\,,
\qquad \gamma_i=\prod_{j\in i}\gamma_j\,.
\end{equation}
The unstable hadronic resonances 
 are allowed to disintegrate and feed the stable hadron spectra.  
Central rapidity region $y\simeq 0$, where 
 $E_i=\sqrt{m_i^2+p^2}=\sqrt{m_i^2+p_\bot^2}\cosh y $\,, 
or the full phase space coverage  as  required  by 
 the kinematic range of the experiments are considered.
Since $p_\bot>1$ GeV hadron spectra are not significantly 
deformed by decays (see Figs. 3 and 4 in Ref.\,\cite{Let95}),
 one may in this limit first evaluate the partial multiplicities
for different hadrons, and than allow these to 
decay, which  speeds considerably the calculations. We have verified that 
this approach, which is exact  for $4\pi$  multiplicities, gives 
results which are much more precise than the experimental 
uncertainties for the central $y\simeq 0$ high $p_\bot$ data, 
and thus we have used this approach.
Once the parameters  $T_{\rm f},\,\lambda_{\rm q},\,\lambda_{\rm s},\,
\gamma_{\rm q},\,\gamma_{\rm s}$ are determined from the   
particle yields available, we can reconstitute the 
entire hadronic particle phase space and obtain the 
physical properties of the system, such as, {\it e.g.}, 
energy and entropy per baryon, strangeness content. 
Even though we are describing a free streaming gas of emitted particles, we 
can proceed as if we were evaluating partition function of system with the 
phase space distributions described 
by the statistical parameters, given that
in a gedanken experiment we have just in an earlier moment still a cohesive,
interacting system. We have implemented all relevant hadronic states and 
resonances in this approach and have also included quantum statistical 
corrections, allowing for first
Bose and Fermi distribution corrections in the hadron abundances and in
the phase space content. These corrections influence favorably the quality 
of the data fits.

Most of our analysis will be carried out within this 
thermo-chemical framework.  However, in section~\ref{secsum} 
we shall enlarge the discussion to include the  analysis of 
$m_\bot$ particle spectra, introducing the collective  velocity 
$\vec v_{\rm c}$.  The different schemes that are possible to 
implement flow were studied  \cite{Hei92}.  We adopt here a radial 
expansion model and  consider the causally disconnected domains 
of the dense matter fireball  to be synchronized at the 
instance of collision --- in other words the time of freeze-out 
is for all volume elements a common constant time in the CM frame. 
The freeze-out occurs at the surface of 
the fireball simultaneously in the CM frame, but not necessarily 
within a short instant of CM-time. 

Within this approach  the spectra and thus also multiplicities 
of particles emitted are obtained replacing 
the Boltzmann factor in Eq.\,(\ref{abund}) by \cite{Hei92}:
\begin{equation}\label{abundflow}
e^{-E_j/T}\to \frac1{2\pi}\int d\Omega_v
  \gamma_v(1+\vec v_{\rm c}\cdot \vec p_j/E_j)
  e^{-{{\gamma_vE_j}\over T}
    \left(1+\vec v_{\rm c}\cdot \vec p_j/E_j\right)} \,,\qquad 
\gamma_v=\frac1{\sqrt{1-\vec v_{\rm c}^{\,2}}}\,,
\end{equation}
a result which can  be intuitively obtained by a Lorentz
transformation between an observer on the surface of
the fireball, and one at rest in laboratory frame. In 
certain details the results we obtain
confirm the applicability  of this simple approach.  

As can be seen in Eq.\,(\ref{abundflow}) we need to carry 
out an additional two dimensional  (half-sphere)  surface 
integral with the coordinate system fixed by an arbitrary, but fixed 
(collision) axis which defines the transverse particle momentum.
Just a one dimensional numerical integration over one of the surface 
angles needs to be carried out. To obtain the particle spectra as function 
of $m_\bot$ we need also to integrate over rapidity $y$\,. This rapidity
integration can be approximated for  a narrow rapidity interval 
using  the error function. 

\section{Analysis of experimental results}
We now discuss the results of our data analysis. In table~\ref{fitsw}, 
we present 10 different least square fits to the data 
obtained using MINUIT96.03 program from the CERN Fortran library. The
two fits (A,\,A$^\prime$) are carried out allowing just two parameters,
$T_{\rm f}$ and $\lambda_{\rm q}$, leaving the remaining three chemical 
parameters at their implicit value ($=1$). The difference between
the two fits is that A (and B,\,C,\,D) exclude data comprising the 
$\Omega$-yields for reasons which we will discuss
momentarily. We consider here 18 data points listed in table~\ref{resultsw} 
(of which three comprise  the $\Omega$'s) which we fit with these
two parameters, thus in fit A$^\prime$ we have 16 and in fit A 13 degrees
of freedom (dof), while in the all-parameter fits D 
and D$^\prime$ we have 13 and  respectively 
10 dof. As seen by the large $\chi^2$/dof shown in last column
of table~\ref{fitsw}, the fit A is not acceptable. Moreover, fitting 
$\lambda_{\rm s}$ (fits B,\,B$^\prime$), and thus in effect 
testing the so called 
hadronic gas model \cite{SCRS93}, is actually not improving the 
validity of the description, even though we have not enforced here
the strangeness conservation among the emitted hadrons. 
Setting $\langle s-\bar s\rangle=0$ 
introduces a constraint between the 3 parameters which is difficult 
to satisfy.  Implementing such strangeness conservation (not shown in 
table~\ref{fitsw}) we find that the fit to all
data has $\chi^2/{\rm dof}\simeq 24$\,, with 16 dof. 
\begin{table}[tb]
\caption{\label{fitsw}
Statistical parameters obtained from  fits of S--Au/W/Pb data.
In fits A to D, particle abundance ratios comprising $\Omega$
are not fitted. In fits A$^\prime$ to D$^\prime$ all 
experimental data in table~\protect\ref{resultsw} 
were used;
in fits D$_{\rm s}$ and D$^\prime_{\rm s}$ strangeness conservation in 
the fitted particle  yields was enforced. Asterisk ($^*$)  
means a fixed (input) value, not a parameter of the fit.}
\vspace{-0.2cm}\begin{center}
\begin{tabular}{lcccccc}
\hline\hline
Fits&$T_{\rm f}$ [MeV]& $\lambda_{\rm q}$&$\lambda_{\rm s}$&
$\gamma_{\rm s}$&$\gamma_{\rm q}$& $\chi^2/$dof \\
\hline
                    A$^\prime$
		     &  145 $\pm$ 3
                 & 1.52 $\pm$ 0.02
                 &   1$^*$
                 &   1$^*$
                 &   1$^*$
                 &   17  \\
                    A
		     &  145 $\pm$ 3
                 & 1.52 $\pm$ 0.02
                 &   1$^*$
                 &   1$^*$
                 &   1$^*$
                 &   21  \\
                    B$^\prime$
		     & 144 $\pm$ 2
                 & 1.52 $\pm$ 0.02
                 &   0.97 $\pm$ 0.02 
                 &   1$^*$
                 &   1$^*$
                 &   18  \\
                    B
		     & 144 $\pm$ 2
                 & 1.53 $\pm$ 0.02
                 &   0.97 $\pm$ 0.02 
                 &   1$^*$
                 &   1$^*$
                 &   22  \\
                    C$^\prime$
		     &  147 $\pm$ 2 
                 & 1.48 $\pm$ 0.02
                 &   1.01 $\pm$ 0.02 
                 &   0.62 $\pm$ 0.02 
                 &   1$^*$
                 &   2.4  \\
                    C
		     &  147 $\pm$ 2 
                 & 1.49 $\pm$ 0.02
                 &   1.01 $\pm$ 0.02 
                 &   0.62 $\pm$ 0.02 
                 &   1$^*$
                 &   2.7  \\
                    D$^\prime$
		     &  144 $\pm$ 3 
                 & 1.49 $\pm$ 0.02
                 &   1.00 $\pm$ 0.02 
                 &   0.73 $\pm$ 0.02 
                 &   1.22 $\pm$ 0.06
                 &   0.90  \\
                 \bf   D
		     & \bf 143 $\pm$ 3 
                 &\bf   1.50 $\pm$ 0.02
                 &\bf   1.00 $\pm$ 0.02 
                 &\bf   0.73 $\pm$ 0.02 
                 &\bf   1.22 $\pm$ 0.06
                 &\bf   0.65  \\
                    D$^\prime_s$
		     &  153 $\pm$ 3 
                 & 1.42 $\pm$ 0.02
                 &   1.10 $\pm$ 0.02 
                 &   0.70 $\pm$ 0.02 
                 &   1.26 $\pm$ 0.06
                 &   3.04  \\
                    D$_s$
		     &  153 $\pm$ 3 
                 &   1.42 $\pm$ 0.02
                 &   1.10 $\pm$ 0.02 
                 &   0.70 $\pm$ 0.02 
                 &   1.26 $\pm$ 0.06
                 &   3.47  \\
\hline\hline
\end{tabular}
\end{center}
\end{table}

In short, the  equilibrium hadron gas model of particle production 
fails today to describe the precise experimental particle abundances, 
our finding contradicts strongly earlier expectations \cite{BSWX96}. 
The major  discrepancies, as shown in table~\ref{resultsw} column B,
include $\Xi/\Lambda$ (10 standard deviations (sd) and 7sd), 
$\overline{\Xi}/\bar\Lambda$ (6sd), $K_{\rm s}^0/\bar\Lambda$ (5sd).
Two values of same particle ratios appear here since if a particle ratio is
 available at fixed $p_\bot$ and 
fixed $m_\bot$, both are fitted and the total $\chi^2$ included
in analysis, as  these experimental results correspond to 
different data samples. We did not 
include in our analysis the same data sample as  seen in  
\cite{BSWX96}, but  the number of degrees of freedom  remains 
similar. We have introduced precision central rapidity 
strange baryon and antibaryon results, and
omitted several results (see below) that we could not consider as having 
a straight-forward thermal interpretation. We do not see our 
data selection as  biased, as in the end we 
are retaining a quite diverse sample typically with smallest errors. 
We did not consider:\\
$\bullet$ the NA44 deuteron yields since it is
 difficult to distinguish the direct, thermally produced
fraction from final state interaction  formation (and depletion);\\
$\bullet$ $\eta/\pi^0$ of WA80 as the flavor evolution of the $\eta$ has not
been fully understood within the thermal model;\\
$\bullet$ $\phi/(\rho+\omega)$ of NA34 (Helios 3) as it is obtained 
in forward rapidity domain and thus subject to considerable flow 
influence not studied in detail here;\\
$\bullet$ the early NA44 pion to nucleon ratios were  used in 
\cite{BSWX96}, but these are obtained from non-overlapping 
rapidity regions, and data was not corrected for the weak hyperon 
feed down decay.  However, we considered that it is very important 
to include a similar data sample in the fit, as the pion to nucleon 
ratio is strongly 
temperature dependent (see below Fig.\,\ref{hmb}). However, the 
results from NA35 are available only for S--Ag reactions, a somewhat lighter 
collisions system. We have looked carefully at the systematics of these 
and other results
offered by the NA35/49 collaboration and have determined that the ratio 
 $h^-/(p-\bar p)$  ($h^\pm$ are all positively/negatively charged hadrons)
shows  little variability between the different systems studied. In S--S 
the result is $4.6\pm0.4$, in S--Ag (see table \ref{resultsw}) it is 
$4.3\pm0.3$ and in Pb--Pb it is $4.4\pm0.5$, see \cite{BGS98} for 
data review.  We thus 
adopted  in our fit the central value observed in S--Ag collisions.  
The smallest error we chose is a conservative assumption here, as it 
in principle pushes the overall error of the fit up. 
In passing we also note that  another related variable we 
consider here in difference to  earlier work \cite{Let95}, is  the ratio 
$(h^+-h^-)/(h^++h^-)$.  We  fit now the $4\pi$ data of EMU05,
rather than the central pseudo-rapidity domain value.

\begin{table}[tb]
\caption{\label{resultsw}
Available particle ratios: experimental results for S--W/Pb/Au, 
references for the results in second column, momentum 
or transverse mass cuts 
(in GeV) in third  column, followed by columns showing the
different fits, corresponding to those in table~\protect\ref{fitsw}.
Asterisk~$^*$ means a predicted result (corresponding data is not 
fitted or not available).}
\small
\vspace{-0.2cm}\begin{center}
\begin{tabular}{lclclllllll}
\hline\hline    
 Ratios
& Refs. & Cuts & Data  &Fit A &Fit B  &Fit C  &\bf Fit D  &Fit D$^\prime$ & Fit D$_s$  &Fit D$^\prime_s$\\
\hline
${\Xi}/{\Lambda}$ &\footnotemark[1]&   
$1.2<p_{\bot}<3$   &0.097 $\pm$ 0.006                & 0.162 & 0.157 & 0.105 &\bf 0.099 & 0.100 &0.110 &0.111\\
${\overline{\Xi}}/{\bar\Lambda}$&\footnotemark[1]&  
$1.2<p_{\bot}<3$ &0.23 $\pm$ 0.02                    & 0.36  & 0.38  & 0.23  &\bf 0.22  & 0.22  &0.18 &0.18\\
${\bar\Lambda}/{\Lambda}$ &\footnotemark[1]&
$1.2<p_{\bot}<3$ &0.196 $\pm$ 0.011                  & 0.194 & 0.202 & 0.205 &\bf 0.203 & 0.203 &0.203 &0.203\\
${\overline{\Xi}}/{\Xi}$ &\footnotemark[1]&    
$1.2<p_{\bot}<3$    &0.47 $\pm$ 0.06                 & 0.43  & 0.48  & 0.44  &\bf 0.45  & 0.44  &0.33 &0.33\\
${\overline{\Omega}}/{\Omega}$&\footnotemark[2]&
$p_{\bot}>1.6$&0.57 $\pm$ 0.41                       &1.00$^{*}$&1.18$^{*}$&0.96$^{*}$&\bf 1.01$^{*}$& 0.98&0.55$^{*}$ &0.55 \\
$\Omega+\overline{\Omega}\over\Xi+\bar{\Xi}$&\footnotemark[2]&
{$p_{\bot}>1.6$}&0.80 $\pm$ 0.40                     &0.27$^{*}$&0.27$^{*}$&0.17$^{*}$&\bf 0.16$^{*}$& 0.16 &0.16$^{*}$ &0.16\\
${K^+}/{K^-}$  & \footnotemark[1]& 
{$p_{\bot}>0.9$}         &1.67  $\pm$ 0.15           & 1.96   & 2.06  & 1.78  &\bf  1.82 & 1.80  &1.43 &1.43\\
${K^0_{\rm s}}/\Lambda$  & \footnotemark[3] & 
{$p_{\bot}>1$}      &1.43  $\pm$ 0.10                & 1.51   & 1.56  & 1.64  &\bf  1.41 & 1.41 &1.25 &1.25\\
${K^0_{\rm s}}/\bar{\Lambda}$ & \footnotemark[3]  &
{$p_{\bot}>1$}  &6.45  $\pm$ 0.61                    & 7.85   & 7.79  & 8.02  &\bf  6.96 & 6.96 &6.18 &6.17\\
${K^0_{\rm s}}/\Lambda$  & \footnotemark[1] & 
{$m_{\bot}>1.9$}    &0.22  $\pm$ 0.02                &0.26    & 0.26  & 0.28  &\bf  0.24 & 0.24 &0.24 &0.24\\
${K^0_{\rm s}}/\bar{\Lambda}$ & \footnotemark[1] &
{$m_{\bot}>1.9$}  &0.87  $\pm$ 0.09                  & 1.33   &1.30   & 1.38  &\bf  1.15 & 1.16 &1.20 &1.20\\
${\Xi}/{\Lambda}$  & \footnotemark[1] &
{$m_{\bot}>1.9$}       &0.17 $\pm$ 0.01              & 0.28   & 0.27  & 0.18  &\bf  0.17 & 0.17 &0.18 &0.18 \\
${\overline{\Xi}}/{\bar\Lambda}$  &\footnotemark[1] &
{$m_{\bot}>1.9$}&0.38 $\pm$ 0.04                     & 0.62   & 0.64  & 0.38  &\bf  0.38 & 0.37 &0.30 &0.30\\
$\Omega+\overline{\Omega}\over\Xi+\bar{\Xi}$ &\footnotemark[1]&
{$m_{\bot}>2.3$}&1.7 $\pm$ 0.9                       &0.95$^{*}$&0.98$^{*}$&0.59$^{*}$ &\bf 0.58$^{*}$& 0.58 &0.52$^{*}$ &0.52\\
$p/{\bar p}$ &\footnotemark[4]&
Mid-rapidity &11 $\pm$ 2\ \                          & 11.0   & 11.2  & 10.1  &\bf 10.6  & 10.5 &7.96 &7.96\\
${\bar\Lambda}/{\bar p}$ &\footnotemark[5]& 
4 $\pi$   &1.2 $\pm$ 0.3                             & 2.43   & 2.50  & 1.47  &\bf 1.44  & 1.43 &1.15 &1.15\\
$h^-\over p-\bar p$  & \footnotemark[6]   &       
4 $\pi$   &4.3 $\pm$ 0.3                            & 4.5   & 4.4  & 4.2  &\bf 4.1  & 4.1 &3.6 &3.6\\
$h^+-h^-\over h^++h^-$  & \footnotemark[7]   &  
4 $\pi$   &0.124 $\pm$ 0.014                         & 0.109  & 0.114 & 0.096 &\bf 0.103 & 0.102 &0.092 &0.092\\
\hline\hline
\end{tabular}
\end{center}
\vspace{-0.2cm}
\footnotesize
$^1$ {S.\,Abatzis {\it et al.}, WA85 Collaboration, {\it Heavy Ion Physics} {\bf 4}, 79 (1996).} \\
$^2$ {S.\,Abatzis {\it et al.}, WA85 Collaboration, {\it Phys.\,Lett.}\,B {\bf 347}, 158 (1995).} \\ 
$^3$ {S.\,Abatzis {\it et al.}, WA85 Collaboration, {\it Phys.\,Lett.}\,B {\bf 376}, 251 (1996).} \\ 
$^4$ {I.G.\,Bearden {\it et al.}, NA44 Collaboration, {\it Phys.\,Rev.}\,C {\bf 57}, 837 (1998).}  \\
$^5$ {D.\,R\"ohrich for the NA35 Collaboration, {\it Heavy Ion Physics} {\bf 4}, 71 (1996).} \\ 
$^6$ S--Ag value adopted here: {T.\,Alber {\it et al.}, NA35 Collaboration, 
{\it Eur.\,Phys.\,J.} C {\bf 2}, 643 (1998); [hep-ex/9711001].}\\
$^7$ A. Iyono {\it et al.}, EMU05 Collaboration, {\it Nucl.\,Phys.\,}A {\bf 544},
455c (1992) and  Y. Takahashi {\it et al.}, EMU05 Collaboration, 
private communication.
 \end{table}

\begin{figure}[tb]
\vspace*{-2cm}
\centerline{\hspace*{-0cm}
\psfig{width=11cm,figure=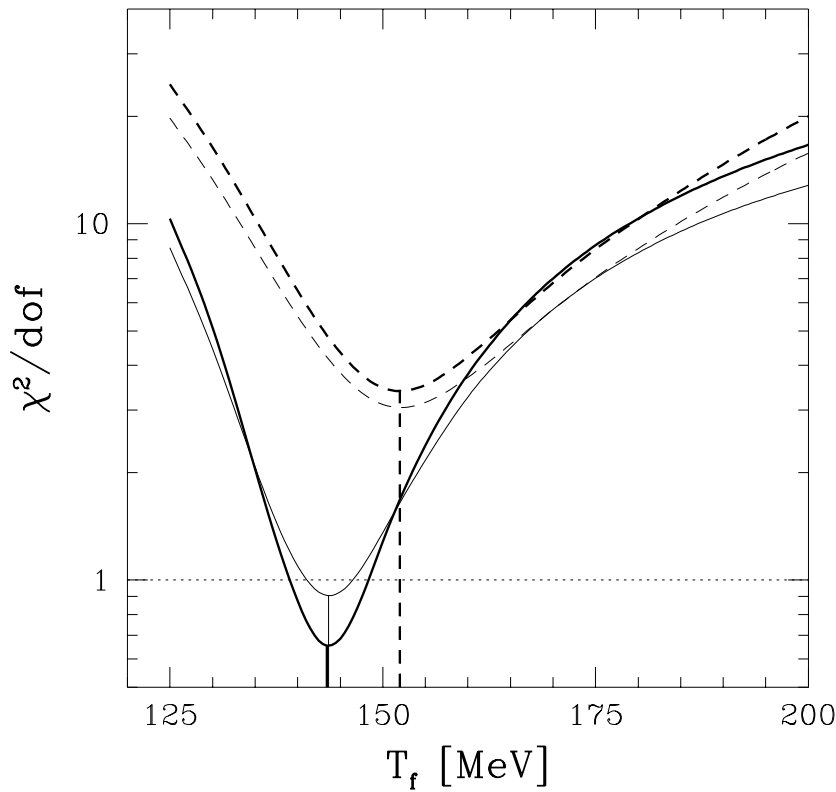}
}
\vspace*{-2.5cm}
\caption{\protect{$\chi^2$}/dof profile as function of temperature
for the chemical non-equilibrium fits: without strangeness conservation
constraint  D (thick solid line), D$^\prime$ (thin solid line), and
with  strangeness conservation constraint D$_{\rm s}$ (thick dashed
line), D$^\prime_{\rm s}$ (thin dashed line).  Thick lines exclude 
the three $\Omega$-data points.  
\label{chi2}
}
\end{figure}
In the fits (C,\,C$^\prime$), we introduce $\gamma_{\rm s}$, {\it i.e.},
strangeness chemical non-equilibrium is
allowed for \cite{Raf91}. This single parameter reduces 
greatly $\chi^2/{\rm dof}$, which drops by a factor 10, 
providing a major improvement in
comparison between theory and experiment. 
The resulting  fit is fully compatible with  
the scenario {\bf B} in \cite{Let95}.  However, presence
of additional and/or more precise experimental data makes this
fit, which was 5 years ago possible,
now statistically also not acceptable. For $18-4=14$ dof 
(fit C$^\prime$ including $\Omega$'s), $\chi^2/{\rm dof}=2.4$ is 
expected to arise in less than 1\% of experiments. Clearly some 
physics is missing in this description of the data. 
It has been  argued that a more refined
dynamical particle production 
model is needed to account for this discrepancy, reflecting on some 
complex dynamics involved in particle production 
(see, {\it e.g.}, \cite{GS98}). 
The surprise is that all that is needed is the
chemical non-equilibrium of the light quark abundance.

We now introduce $\gamma_{\rm q}$.
 The full fit to all data (fit D$^\prime$) is
statistically significant, with $\chi^2$/dof = 0.9\,. The profile of
$\chi^2$/dof as function of $T_{\rm f}$ shown 
in Fig.\,\ref{chi2} (thin solid line) shows a well defined 
minimum at $144\pm3$ MeV. We note, in passing, that quite different 
considerations have lead others to propose a universal 
chemical freeze-out
temperature at nearly exactly this value \cite{NXU}. 
The thick solid line in Fig.\,\ref{chi2}  (fit D) 
arises when, in the fit, one excludes 
three data points involving completely strange $\Omega$'s. We came to
 try this, since when exploring the stability of 
the fit D$^\prime$ against suppression of
some of the data, we noted that the fit D as expressed by $\chi^2$/dof 
is better. 
As can be seen in Fig.\,\ref{chi2}, the fit D minimum (thick solid line
at $T_{\rm f}=143$ MeV) is sharper and $\chi^2$/dof
 drops to 0.65\,. 
In the  table~\ref{fitsw}, we further see that
the removal of these three data points has for all the other fits the effect of
reducing the significance of the fit, as would be expected given the
few degrees of freedom we have at our disposal. 
That an improvement occurs when 
the fit is statistically significant signals that there is probably
an additional mechanism of $\Omega$ formation contributing to the yields. 
This indeed is what
we expect, since we have not enforced the strangeness conservation
in the fits D (in bold face) and D$^\prime$ shown in  
table~\ref{fitsw}. Enforcing
strangeness conservation leads to fits denoted 
(D$_{s}$,\,D$^\prime_{s}$) 
with $\chi^2$/dof $>3$, and thus  not offering an acceptable interpretation
of the data. There indeed is also not a well defined $\chi^2$ minimum 
in the associated distributions shown by dashed lines in  
Fig.\,\ref{chi2}. 

Thus the following simple reaction scenario emerges: 
in the particle evaporation-dissociation 
process at $T_{\rm f}=143\pm3$ MeV there is predominant 
emission of $\bar s$-carrying hadrons (see below) leading to a $s$-carrier 
rich residue that populate lower $p_\bot$ kaon and hyperon yields that
remained not fully measured.  
It should be here stressed that the data we fit are very sparse in
the small $p_\bot<1$ GeV domain, thus an imbalance between $\bar s$
and $s$ we show in table~\ref{tqsw}, simply should be read to mean that 
the opposite imbalance is to be expected for soft hadrons.
A soft hadron excess arising from  disintegration of $s$-quark
enriched residue becomes particularly 
visible in the $\Omega$-abundance, since
the abundance of these particles is  small, while their production by 
strangeness rich source is probably enhanced, speaking 
here in relative terms.
We note that such strangeness distillation has been foreseen 
to occur \cite{GKS87}, and specifically in this manner 
if QGP phase hadronizes
at temperatures  found in our fits of the data \cite{Raf87}. 
This dynamical distillation process stimulated several 
searches for strangeletts, {\it i.e.}, (quasi) stable highly strange
hadrons with more than three strange quarks \cite{E864,NA52}.

\section{Properties of hadronic particle phase space}
Given the precise statistical information about the hadron phase space
provided by the fit D, we can determine the specific 
content in energy, entropy, strangeness carried out 
by hadronic particles, see table~\ref{tqsw}. For each fit we show the 
chemical freeze-out temperature $T_{\rm f}$, the hadron phase space
properties: specific (per baryon) energy, entropy, (anti)strangeness,
strangeness imbalance, and the freeze-out pressure $P_{\rm f}$ and 
volume $V_{\rm f}$.
The specific $\bar s$ content is determined to be $0.90\pm0.04$. 
We find 
that there is  an excess of $\bar s$-carrying hadrons carried away 
at higher $p_\bot$. 
The specific entropy content $48.2\pm3$ of the best fit D,
agrees well with the evaluation made 
earlier \cite{Let93}. This is so because the fit of predominantly high 
$p_\bot$ strange particle data is fully consistent with the $4\pi$
total multiplicity results. The pressure of the hadronic phase space 
$P_{\rm f}=82\pm6\,{\rm MeV/fm}^3$ has the magnitude of the vacuum 
confinement pressure.
If the chemical freeze-out were the result of a disintegration
after QGP expansion to the point of instability into volume 
fragmentation, this would be according to the Gibbs criterion
the transformation  pressure. In that case our fit amounts to the 
measurement of the confinement pressure (bag constant ${\cal B}$). 
The volume $V_{\rm f}$ we quote arises by assuming that the hadronic
phase space comprises for the S--W collisions baryon number $B=120$. 
It implies in the case of fit D a hadron phase space 
energy density $\epsilon_{\rm f}=0.43\pm 0.4 \,{\rm GeV/fm}^3$. 
Both, $V_{\rm f}$ and $\epsilon_{\rm f}$ have physical meaning 
only if the chemical freeze-out occurs nearly simultaneously  
within the entire body of the source.

\begin{table}[tb]
\caption{\label{tqsw}
$T_{\rm f}$ and physical properties of  the full 
hadron phase space characterized by the statistical parameters  
given in table~\protect\ref{fitsw}.} \small
\vspace{-0.2cm}\begin{center}
\begin{tabular}{lccccccc} 
\hline\hline
Fits&$T_{\rm f}$ [MeV]& $E_{\rm f}/B$  & $S_{\rm f}/B$ & ${\bar s}_{\rm f}/B$ & $({\bar s}_{\rm f}-s_{\rm f})/B$ &$P_{\rm f}$ [GeV/fm$^3$] & $V_{\rm f}$ [fm$^3$]\\
\hline
                    A
		     &  145 $\pm$ 3
                 & 9.01 $\pm$ 0.50
                 &  50.1 $\pm$ 3
                 &  1.64 $\pm$ 0.06
                 &  0.37 $\pm$ 0.02 
                 &  0.056 $\pm$ 0.005 
                 &  3352 $\pm$ 350  \\
                    B
		     & 144 $\pm$ 2
                 & 8.89 $\pm$ 0.50
                 &  50.0 $\pm$ 3
                 &  1.66 $\pm$ 0.06
                 &  0.44 $\pm$ 0.02  
                 &  0.056 $\pm$ 0.005  
                 &  3343 $\pm$ 350  \\
                    C
		     &  147 $\pm$ 2
                 & 9.25 $\pm$ 0.50
                 & 48.5 $\pm$ 3
                 & 1.05 $\pm$ 0.05
                 & 0.23 $\pm$ 0.02  
                 & 0.059 $\pm$ 0.005  
                 & 3529 $\pm$ 350  \\
                 \bf   D
		     &\bf   143 $\pm$ 3
                 &\bf  9.05 $\pm$ 0.50
                 &\bf   48.2 $\pm$ 3
                 &\bf   0.91 $\pm$ 0.04
                 &\bf   0.20 $\pm$ 0.02  
                 &\bf   0.082 $\pm$ 0.006  
                 &\bf   2524 $\pm$ 250  \\ 
                    D$^\prime$
		     &  144 $\pm$ 2
                 & 9.07 $\pm$ 0.50
                 & 48.3 $\pm$ 3
                 & 0.91 $\pm$ 0.05
                 & 0.20 $\pm$ 0.02  
                 & 0.082 $\pm$ 0.006  
                 & 2521 $\pm$ 250  \\
                    D$_s$
		     &   153 $\pm$ 3
                 &  8.89 $\pm$ 0.50
                 &  45.1 $\pm$ 3
                 &  0.76 $\pm$ 0.04
                 &  0$^*$  
                 &  0.133 $\pm$ 0.008
                 &  1520 $\pm$ 150  \\ 
                    D$^\prime_s$
		     &  153 $\pm$ 3
                 & 8.87 $\pm$ 0.50
                 & 45.1 $\pm$ 3
                 & 0.76 $\pm$ 0.05
                 & 0$^*$  
                 &  0.134 $\pm$ 0.008
                 &  1515 $\pm$  150 \\
\hline\hline
\end{tabular}
\end{center}
\end{table}

\begin{figure}[tb]
\vspace*{0cm}
\centerline{\hspace*{1.5cm}
\psfig{width=12cm,figure=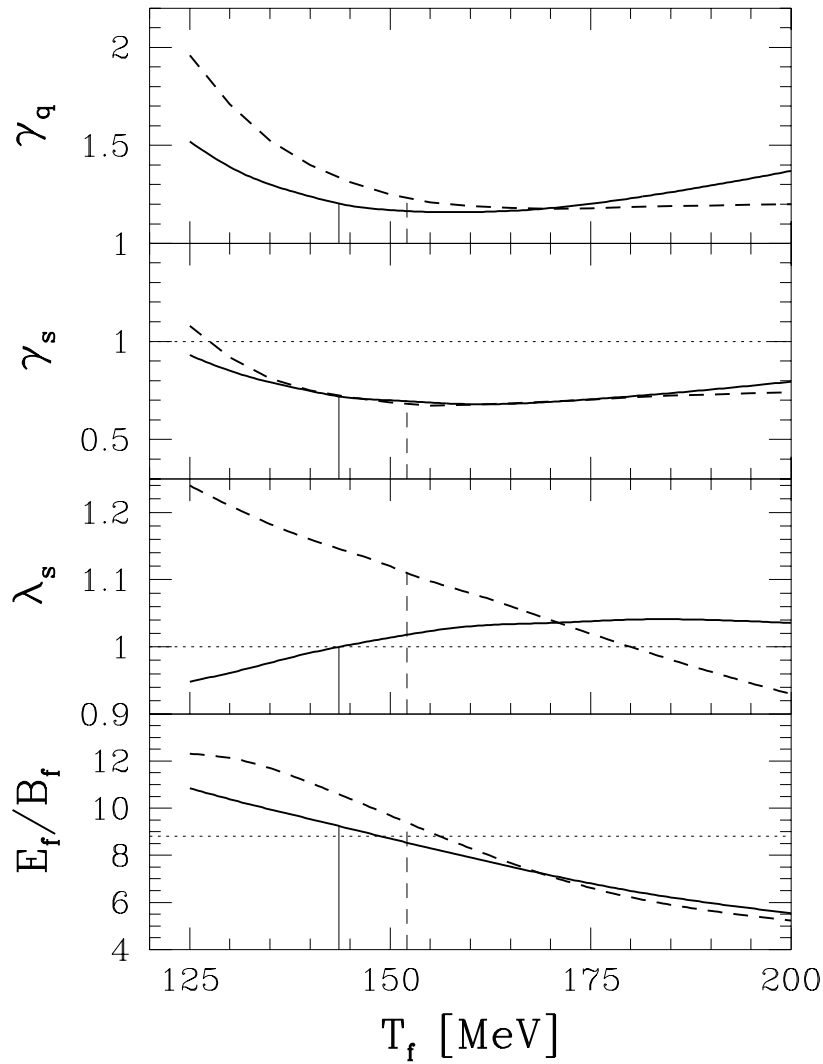}
}
\vspace*{-2.3cm}
\caption{Variation of $E_{\rm f}/B_{\rm f}$, $\lambda_{\rm s}$, 
$\gamma_{\rm s}$, and $\gamma_{\rm q}$ 
as function of temperature $T_{\rm f}$, with all other parameters fixed by 
least square fit to the data.   
\label{Tdep}
}
\end{figure}

The energy per baryon found in emitted hadronic particles
is remarkably consistent with the expectation based on kinematic
energy content, should the deposition of energy and baryon number 
in the fireball be identical \cite{LTR94}. For the favored fit D,
we find $E_{\rm f}/B=9.05\pm0.5$ GeV, to be compared with
kinematic value  8.8 GeV obtained for central collision of Sulphur with 
a tube of matter in the larger target. Moreover, the different 
fits cross the kinematic value (dotted  horizontal line in Fig.\,\ref{Tdep}, 
bottom portion) very near to best fit temperatures.  
The computed values of $E_{\rm f}/B$ are 
shown in Fig.\,\ref{Tdep} for a given  $T_{\rm f}$ for fit D (solid line, 
no $\Omega$ or strangeness conservation) and fit D$_{\rm s}$
(dashed line, no $\Omega$, with   strangeness conservation), with 
all the other parameters obtained from a least square fit at given 
temperature. The locations of the best $\chi^2$/dof are 
indicated by vertical lines. In second 
section from the bottom in  Fig.\,\ref{Tdep}, we draw attention to
another remarkable physical result of the fit D:
the value $\lambda_{\rm s}=1.00\pm0.02$ suggests as
 the source of these particles a state symmetric between
$s$ and $\bar s$ quarks, naively expected in the deconfined
phase. It is not an accident, and 
we could have easily found another result, since $\lambda_{\rm s}$
varies as function of fitted $T_{\rm f}$.

The non-equilibrium parameters, shown in top two sections of 
Fig.\,\ref{Tdep}, are opening the minimum in the $\chi^2$/dof
 distribution  shown in Fig.\,\ref{chi2},
 without a major dislocation in the parameter space 
as is seen in table~\ref{fitsw}. 
The result of interest here is that both consistently differ 
from their equilibrium values $\gamma_{\rm s}=1$, $\gamma_{\rm q}=1$\,.
The values $\gamma_{\rm s}=0.73\pm0.02$ and 
 $\gamma_{\rm q}=1.22\pm0.06$ are compatible with the physics based 
expectations: the computed rate of strangeness production in the
deconfined phase is suggestive of this result
which is noticeably below the full phase space equilibration
 \cite{acta96,RM82}. 
Fragmentation  of
remaining gluons at $T_{\rm f}=143$ MeV naturally leads  to the
value of $\gamma_{\rm q}$ we obtain. As shown in the top half of  
Fig.\,\ref{Tdep}, for the entire range of freeze-out temperature of 
physical interest,  the values of $\gamma_{\rm s}$ and $\gamma_{\rm q}$ 
vary somewhat, but consistently remain in the physical domain 
described above, and are similar for fits with 
and without strangeness conservation. This systematic clearly 
assures that these parameters are here to stay. Even a short view of 
Fig.\,\ref{Tdep} shows that strangeness conservation condition,
which introduces the greatest systematic uncertainty into this
study, is capable to influence the value of $\lambda_s$, but not 
the chemical non-equilibrium parameters shown 
in table~\ref{fitsw}, or other interesting properties of the
hadron source displayed in table~\ref{tqsw}. 

\begin{figure}[tb]
\vspace*{-2.cm}
\centerline{\hspace*{0cm}
\psfig{width=10cm,figure=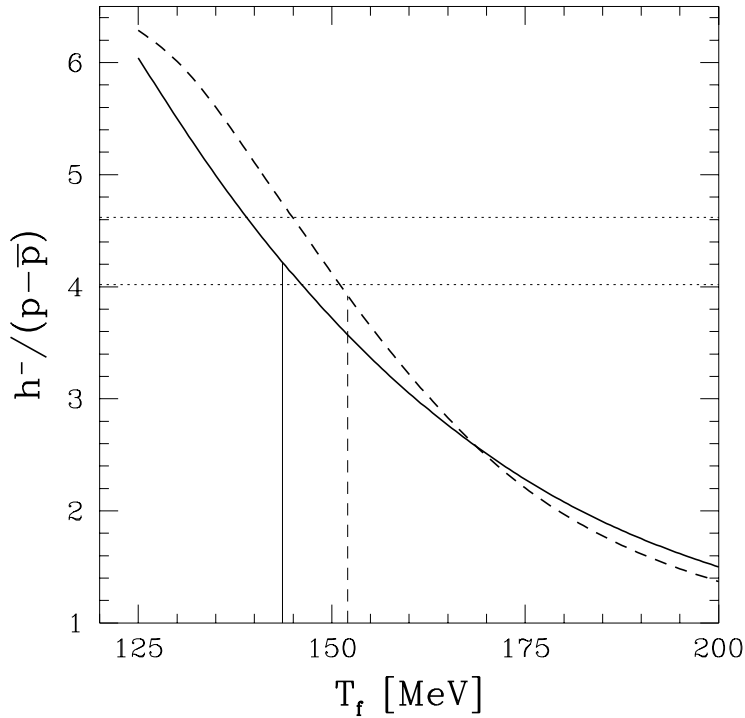}
}
\vspace*{-2.5cm}
\caption{Variation of $h^-/(p-\bar p)$ as function of $T_{\rm f}$ for fits  D
(solid line) and strangeness conserving  D$_{\rm s}$ (dashed line). 
Vertical lines point to best fit location, horizontal dotted lines
brace the experimental result (see table~\ref{resultsw}).
\label{hmb}
}
\end{figure}

The sensitivity of the computed  specific negative hadron yield to the 
freeze-out temperature $T_{\rm f}$ is shown  in Fig.\,\ref{hmb}. For a given 
value of $T_{\rm f}$  all the other parameters follow values determined by least 
square fit. The solid line is
for fit D, dashed line for strangeness conserving fit D$_{\rm s}$. 
Vertical lines are indicating the location of the best fits, horizontal 
dotted lines show the range of the NA35-experimental result for S--Ag 
collisions. The result shown
in Fig.\,\ref{hmb} makes it clear that it is impossible to consider chemical
freeze-out to occur outside of the temperature window $135<T_{\rm f}<155$\,MeV.
Our use of all accessible experimental particle production data 
in S--Au/W/Pb reactions  eliminates thus from further consideration
the recent suggestion of hadronic particle production universality
between $e^++e^-,\,p+\bar p,\, p+p,\, A+A$ reactions with chemical
 freeze-out at $T_f\simeq 180$\,MeV \cite{BGS98}. In our view,
it is more natural  to see earlier ({\it i.e.}, higher $T_{\rm f}$) 
chemical freeze-out in light collision systems, as compared to
heavy ion collisions. We note that in \cite{BGS98}, as here, the 
chemical freeze-out conditions are studied, and that subsequent to
establishment of particle abundances there could still be particle number
conserving (elastic) evolution of the hadronic matter to the point of full
decoupling at a lower temperature. During this evolution the 
final shape of the particle spectra is determined. Thus data 
fits based on details of particle spectra, in particular including
the low $m_\bot$, are geared to yield a thermal freeze-out temperature 
$T_{\rm th}$ that is yet lower than our $T_{\rm f}$ \cite{Nix98}. 

\section{Impact of collective flow on freeze-out properties}\label{secsum}
So far we have considered only  the particle yields, and have
not addressed particle spectra which, as mentioned in beginning of
section \ref{secnoneq}, requires that we allow for the Doppler-like
blue shift of the particles emitted from a moving source at hadronization.
To do this we need to explicitly add to the least square fit
another parameter, the collective  velocity $\vec v_{\rm c}$. 
While the integral over  the entire phase space of the flow spectrum 
yields as many particles with and without flow, when acceptance cuts are 
present particles of different mass experience differing flow effects. 
We consider here the radial flow model, perhaps of the  simplest of
the complex flow cases possible \cite{Hei92}, 
but it suffices to fully assess the impact of flow on our analysis.

For given pair of values $T_{\rm f}$ and $v_{\rm c}$, the resulting  
$m_\bot$ particle spectrum is obtained and analyzed using the shape 
and procedure employed by the experimental groups, and 
the inverse slope `temperature' $T_{\rm s}^j$ is determined 
for each particle $j$ --- since we have the experimental 
results for $T_{\rm s}^j$\,, we can indeed include these in 
our data fit. $v_{\rm c}$ is thus just one additional parameter 
and there are in principle several new data points available.
However, in consideration of their similarity \cite{Eva96}, with
values overlapping  within error, 
we decided to include in the chemical fit only one  value
$T_{\rm s}=235\pm10$, chosen near to the most 
precise Lambda spectra. Once the chemical fit yields $T_{\rm f}$
and $v_{\rm c}$, we check how other particles have fared. The resulting 
$T_{\rm s}^j$ are  in remarkable agreement with experiment, well beyond
what was fitted: we find for kaons, lambdas  and cascades the 
values $T_{\rm s}^j=215, 236$ and 246 MeV respectively,
which both in trend and value agrees with the K$^0$, 
$\Lambda,\,\overline\Lambda$, $\Xi$ and $\overline\Xi$ WA85 
results~\cite{Eva96}: $T_{\rm s}^{{\rm K}^0}=219\pm5,\ 
T_{\rm s}^\Lambda=233\pm3,\
T_{\rm s}^{\overline\Lambda}=232\pm7\,,\ T_{\rm s}^\Xi=244\pm12$ and 
$T_{\rm s}^{\overline\Xi}=238\pm16$.  

Since the flow effect shifts particles of different mass differently
into different domains of $m_\bot,\,y$\,, it is not surprising that the
inclusion of flow only impacts the phase space abundance parameters,
beyond the errors of the fits. The fit D remains the best, it has now
$\chi^2/$dof=0.73, and the parameter values we gave in abstract. Fit D' 
has $\chi^2/$dof=0.83\,. In order to facilitate 
comparison with other work we note that to greater precision our 
fit D with flow yields $T=142.7\pm2.1$ MeV, $\mu_b=176\pm3$ and 
$v_{\rm c}=0.486\pm0.010\,c$. The situation with strangeness
non-conservation remains unchanged, both fits D$_s$ and D$_s$' have 
negligible confidence level with $\chi^2$/dof$\simeq 3.2$. The 
strangeness imbalance in fits D, D' is the same as we have 
obtained without flow. Thus this imbalance is not result of flow
effects, but is an intrinsic property of the evolution of the 
fireball. An interesting feature of with flow fit is that there 
is little  correlation between now 6 parameters in the fit, in other
words the flow velocity is a truly new degree of freedom in description
of the experimental data. We checked that  nearly the same flow 
velocity is found when we disregard in the fit  the experimental 
inverse slope. It is for this reason that we  present here the
simple radial flow model, as within this scheme the inverse transverse
slope of hadrons is correctly `predicted' by the chemical freeze-out 
analysis with flow. We note that in just one aspect the fits with flow 
offer a new insight:  the value of $\gamma_s$ we obtained
is compatible with  unity. This value was noted already in the 
fit  of the S--S collision results \cite{SGHR94}, and so far eluded 
the analysis of S--W/Au/Pb  collisions. 

With   $\gamma_q^2\simeq 2$  the analysis of excess entropy production
\cite{Let93} is fully confirmed, and indeed in fit D
with flow the entropy per baryon is 46.3\,, within the bounds shown
in table \ref{tqsw}. The energy content in the phase space is 
slightly different: the intrinsic  energy per baryon at freeze-out
is $E/B=8.55$ GeV, which when multiplied with the
factor $\gamma_v$ reaches $E/B=9.78$  GeV. 
The freeze-out pressure $P_{\rm f}=0.123$ GeV/fm$^3$, roughly growing  
with a power of $\gamma_q$, has accordingly increased.

As this discussion has shown, the analysis of particle abundances
 and spectra has to be combined in order to be certain that all
the physical parameters acquire their true value. Even though
there is still considerable uncertainty about other freeze-out 
flow effects, such as longitudinal
flow (memory of the collision axis), the level of consistency
and quality of our fit suggests that for the observables considered 
here these effects do not matter. Considering the quality of 
the data description obtained using the chemical non-equilibrium 
parameters,  it is impossible to classify  results presented here
as accidental and likely to see further major revision. 

In conclusion, we have here shown that the hard ($p_\bot> 1$ GeV)
strange particle production data, combined with the global 
hadron multiplicity (entropy), can be consistently 
interpreted within a picture of a hot hadronizing blob of matter
governed by statistical parameters acquiring values 
expected if the source structure
is that of deconfined QGP. We have shown that radial flow allows
to account exactly for the difference between freeze-out temperature 
and the observed spectral shape, and allows full description
of inverse slope of $m_\bot$ strange baryon and antibaryon spectra. 
We find  a highly significant fit to
data with a source characterized by 
$\lambda_{\rm s}=1$, specific baryon energy $E/B\simeq 9$ GeV,
high specific entropy $S/B\simeq 50$, and chemical
freeze-out temperature $T_{\rm f} = 143$ MeV . The dense  blob of 
matter was  expanding with surface 
velocity $v_{\rm c}\simeq 0.49\,c$. The near equilibrium abundance of 
strange quarks ($\gamma_{\rm s}\simeq 1$, including flow), and the
 over-abundance of light quarks ($\gamma_{\rm q}^2\simeq 2$), is 
pointing to a deconfined, fragmenting quark-gluon fireball as the
direct particle emission source. At freeze-out we find the particle 
phase space pressure $P_{\rm f}\simeq 0.1\,{\rm GeV/fm}^3$, which
maybe interpreted as the pressure needed to overcome the external 
vacuum pressure  (bag constant ${\cal B}$). Strangeness non-conservation 
at high $p_\bot$ seen in our fit favors an evolution scenario in which the 
process of  hadronization into hard particles enriches by distillation 
the fireball remnant with strangeness content. Given an excess of 
$\bar s$ emission at high $p_\bot$ of 0.2 strange particles per baryon, 
we expect at  small $p_\bot$ a non-negligible 20\% excess of 
$s$-carriers over $\bar s$-carriers, potentially helping the formation 
of strange quark matter nuggets.  The lessons learned in this study  
will likely prove  helpful in the forthcoming interpretation of the
Pb--Pb data, and the future RHIC-BNL data analysis. 

{\vspace{0.5cm}\noindent\it Acknowledgments:\\}
This work was supported in part 
by a grant from the U.S. Department of
Energy,  DE-FG03-95ER40937\,. \\LPTHE-Univ. Paris 6 et 7 is: 
Unit\'e mixte de Recherche du CNRS, UMR7589.


\end{document}